# Study of magnetic alloy cores for HIRFL-CSRm compressor cavity*

MEI Li-Rong(梅立荣)[1,2;1)] XU Zhe(许哲)[1] YUAN You-Jin（原有进）[1] JIN Peng（金鹏）[1] BIAN Zhi-Bin（边志彬）[1] ZHAOHong-Wei（赵红卫）[1]

1（Institute of Modern Physics, Chinese Academy of Sciences, Lanzhou 730000,China）

2 (Graduate University of Chinese Academy of Sciences, Beijing 100049, China)

**Abstract:** For selecting the properly magnetic alloy (MA) material to load the RF compression cavity, the measurement of the MA cores which is produced by Liyuan Company has been carried out at IMP. We measured 4 kinds of MA core materials, type V1, V2, A1 and A2. And we mainly focus on the permeability, quality factor (Q value) and shunt impedance of the MA core. The MA cores which have higher permeability, lower Q value and higher shunt impedance will be selected to load RF compression cavity. According to the results of measurement, the type V1, V2 and A2 material will be chosen as candidate to load RF cavity.

## 1 Introduction

HIRFL (Heavy Ion Research Facility in Lanzhou) can provide high energy heavy ion beams after HIRFL-CSR (heavy ion cooling storage ring) put into operation. HIRFL-CSR consists of a main ring (CSRm) and an experiment ring (CSRe). For the purpose of actualizing high energy density physics and plasma physics research at HIRFL-CSR, the higher accelerating voltage is required to compress the beam pulse in the longitudinal direction so that the energy of high current intensity and short pulse heavy ion beam can deposit in the experimental target material massively and effectively with a pulse length as short as possible [1]. A magnetic alloy (MA)-loaded RF cavity is a device which can provide higher accelerating voltage to meet the requirement [2]. In addition the MA-loaded RF cavity does not need the tuning loop, so, it simplifies the whole RF control system, and it can be used in the compact accelerator which will be constructed as the cancer therapy facility in the future. In this system the magnetic alloy material plays a very important role. Its characteristics decide the performance of cavity to some extend. So, for selecting the properly magnetic alloy material to load the RF compression cavity, the measurement of the MA cores which is produced by Liyuan Company has been carried out at IMP. We mainly measured the permeability, Q value and shunt impedance of the MA core. Finally the MA cores which have higher permeability, lower Q value and higher shunt impedance will be selected to load RF compression cavity.

*Supported by Accelerator Innovation Community (10921504)

1) Email: lrmei@impcas.ac.cn

2. Permeability of MA cores

The series complex permeability of MA cores can be expressed as

$$\mu = \mu_s' - j\mu_s'' , \tag{1}$$

The real part $\mu_s'$ represents the reserved energy in the process of magnetization; imaginary part $\mu_s''$ represents the dissipation energy. So the quality factor of the MA core can be expressed as

$$Q = \frac{\mu_s'}{\mu_s''} . \tag{2}$$

On the other hand, the permeability of MA cores also can be expressed as the parallel expression

$$\frac{1}{\mu} = \frac{1}{\mu_P'} - \frac{1}{i\mu_P''} , \quad Q = \frac{\mu_p''}{\mu_p'} . \tag{3}$$

So a relation can be obtained according to simplify above expressions.

$$\mu_p' = (\frac{1}{Q^2} + 1)\mu_s' , \quad \mu_p'' = (1 + Q^2)\mu_s'' . \tag{4}$$

This is relationship of the parallel permeability and the series permeability. It will be used in the later section. [3-7]

3. MA cores measurements

3.1 Measurement scheme

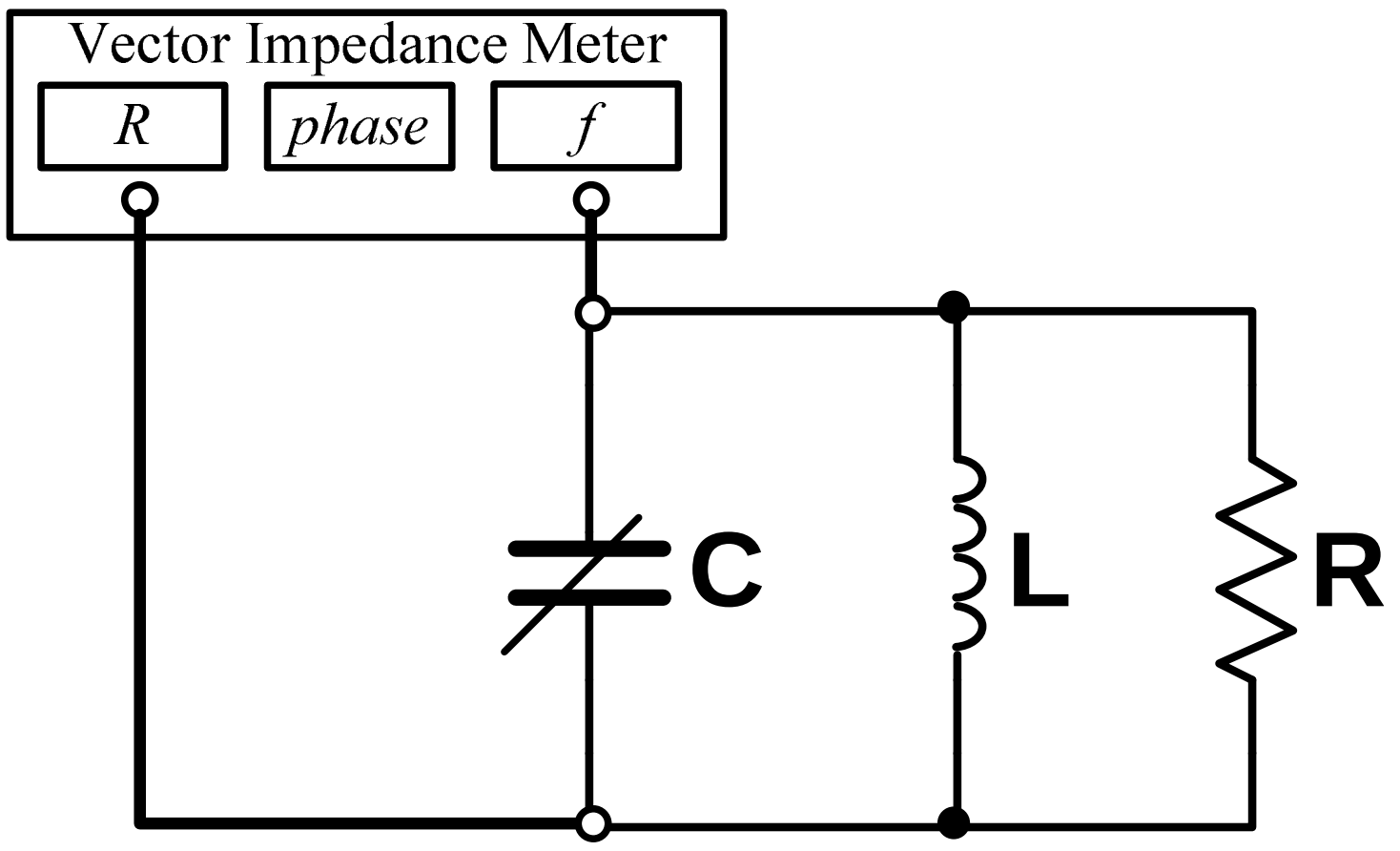

Fig.1. Measurement scheme equivalent circuit

The measurement scheme equivalent circuit of MA core characteristics is shown in Fig. 1. The MA core winding with coil can form an inductance. A parallel resonant circuit is formed when we parallel connect a capacitor C on the inductance. Measure the resonant circuit using vector impedance meter. The resonant frequency $f_0$ and resonant impedance $R_s$ can be obtained. So the inductance of the resonant circuit can be calculated by the following equation,

$$L = \frac{1}{4\pi^2 f_0{}^2 C}. \tag{5}$$

In addition, for MA core the inductance also can be calculated by

$$L = \frac{\mu_0 \mu'_p}{2\pi} \ln \frac{r_2}{r_1} d, \tag{6}$$

So according to above two equations we can get

$$\mu'_p = \frac{1}{2\pi\mu_0 f_0{}^2 C \ln \frac{r_2}{r_1} d} \tag{7}$$

$r_1$, $r_2$ and $d$ are inner radius, outer radius and thickness of the MA core respectively, $\mu_0 = 4\pi\times10^{-7}$ H/m is the permeability of vacuum, $\mu_p'$ is the real part of the parallel complex permeability of MA core.

For the parallel circuit the quality factor Q is calculated by

$$Q = 2\pi f_0 R_s C, \tag{8}$$

According to expression (2) and (4) the $\mu_s'$, $\mu_s''$ can be calculated; finally the permeability of MA cores can be gotten. By changing the value of capacitor which parallel connect with MA core, resonant frequency dependence of the permeability of MA core can be obtained. [8]

3.2 Results of measurement

We measured 4 kinds of MA core materials. Type V1, V2 which outer radius, inner radius and thickness are 65mm, 35mm, and 30mm respectively. Type A1, A2 with dimension of 65mm, 35mm and 25mm. The difference between V1 and V2, A1 and A2 is the insulation layer. The insulation layer of A1 and V1 are the SiO2 and silica gel, while the insulation layer of A2 and V2 is silica gel. In the process of measurement we wind one coil on the MA core. Firstly we measured the impedance magnitude and phase of MA core at 1.0MHz which is the center frequency of frequency range from 0.8 to 1.2MHz. The purpose is to test the impedance and phase change of the MA core when frequency varied. The frequency dependence of impedance magnitude and phase of type V1, V2, A1, A2 core are shown in figure 2.

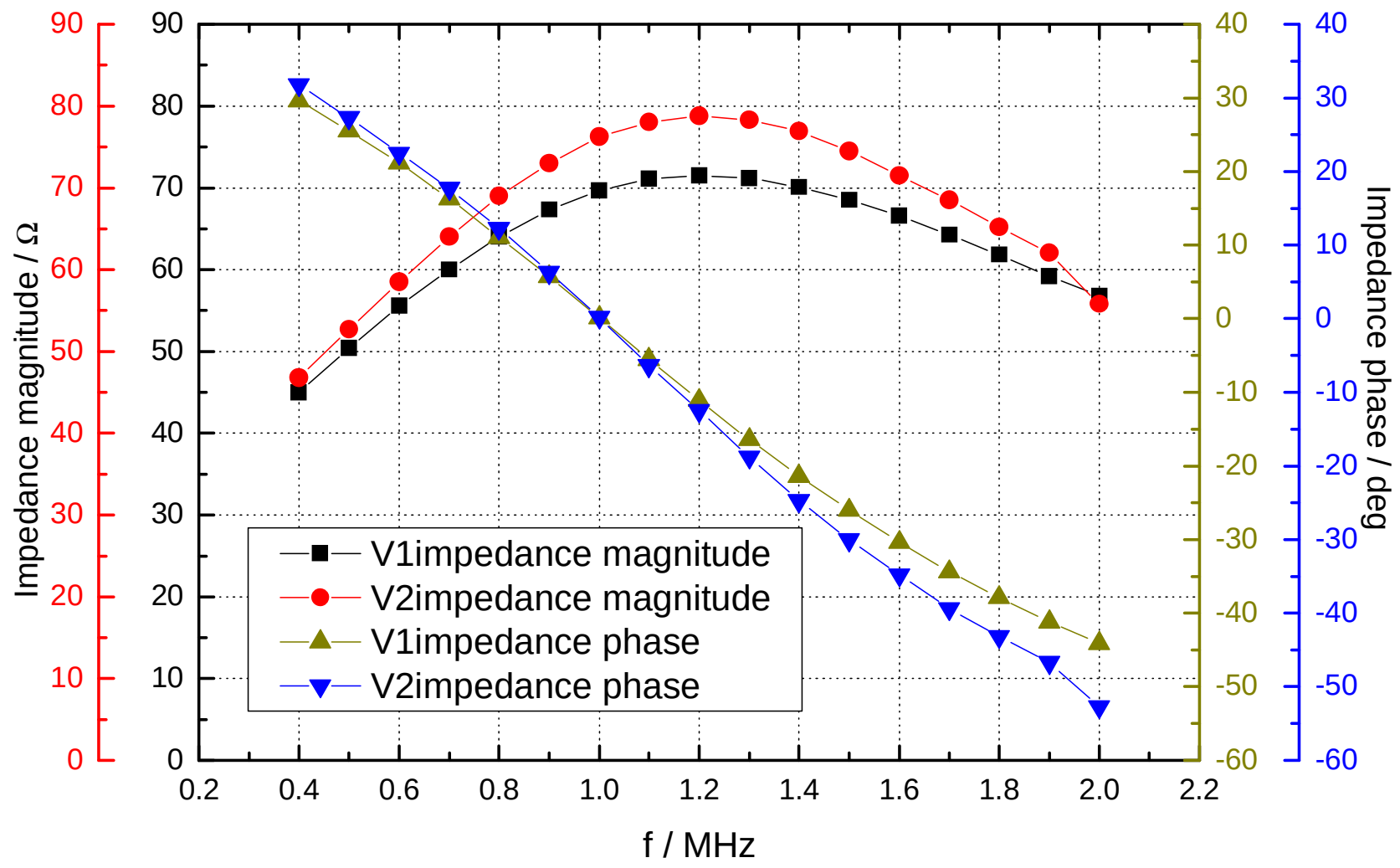

(a). Frequency dependence of impedance and phase of V1 and V2.

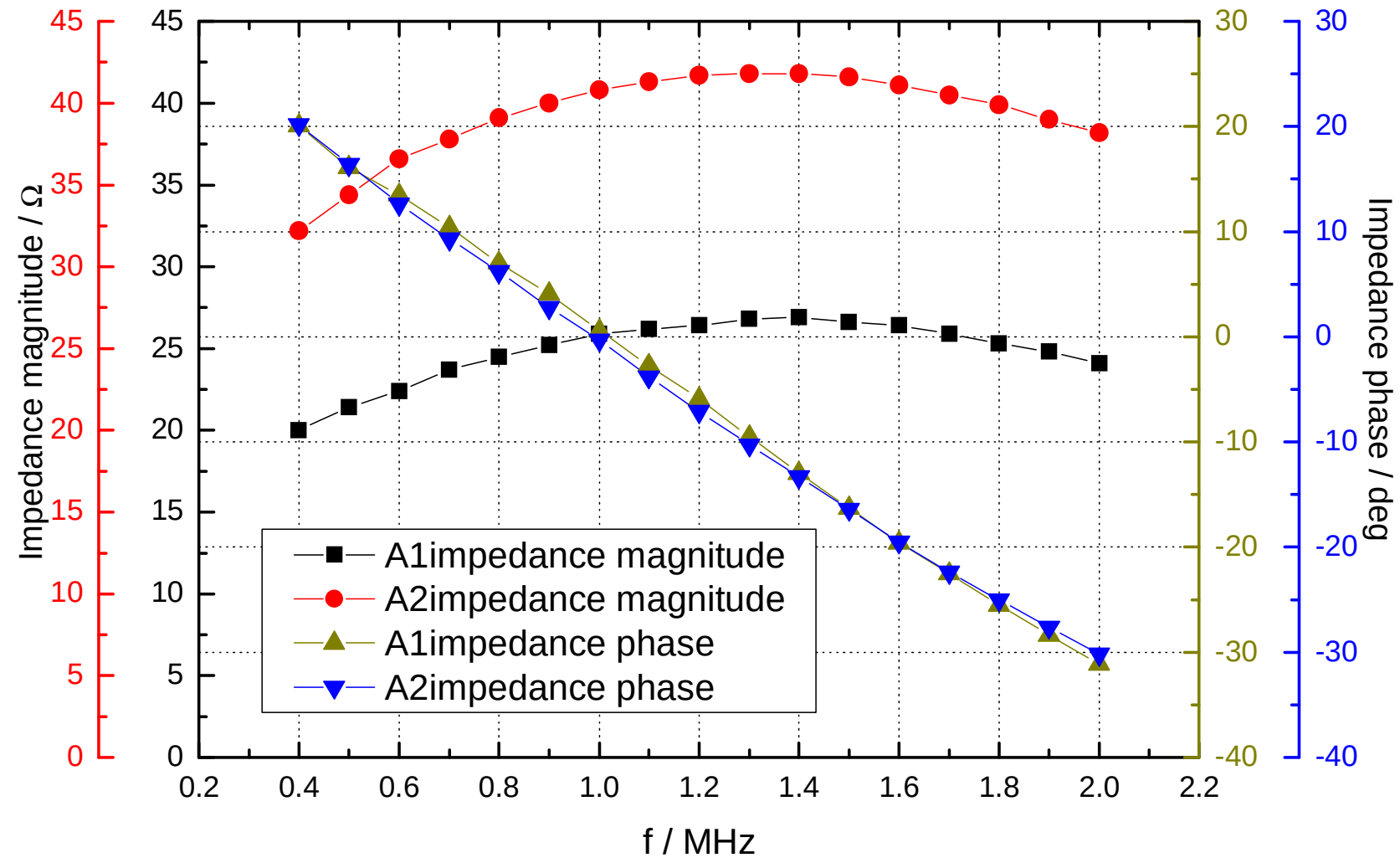

(b). Frequency dependence of impedance and phase of A1 and A2.

Fig.2. Frequency dependence of impedance magnitude and phase of type V1 and V2 (a), A1 and A2 (b) core

From Fig.2 we see that, the impedance of type V1 material is about 70 Ω at 1.0MHz, and the impedance phase varied from 11.1 degree to -11.1 degree when frequency varied from 0.8MHz to 1.2MHz. We also can see the impedance of type V2, A1 and A2 material are about 76.3 Ω, 26 Ω and 41 Ω at 1.0MHz respectively, and the corresponding impedance phase varied from 12.3 to -12.5 degree, 7 to -5.9 degree and 6.2 to -7.1 degree respectively when frequency varied from 0.8MHz to 1.2MHz. These results indicated the impedance change is very small when frequency varied from 0.8MHz to 1.2MHz. This is a better case for an untuned RF cavity.

In addition, we measured the Q value, $\mu_p$'Qf product, $\mu_s$' and shunt impedance of the core at different frequency by changing capacitor. The measured results are shown in

Figs. 3-5

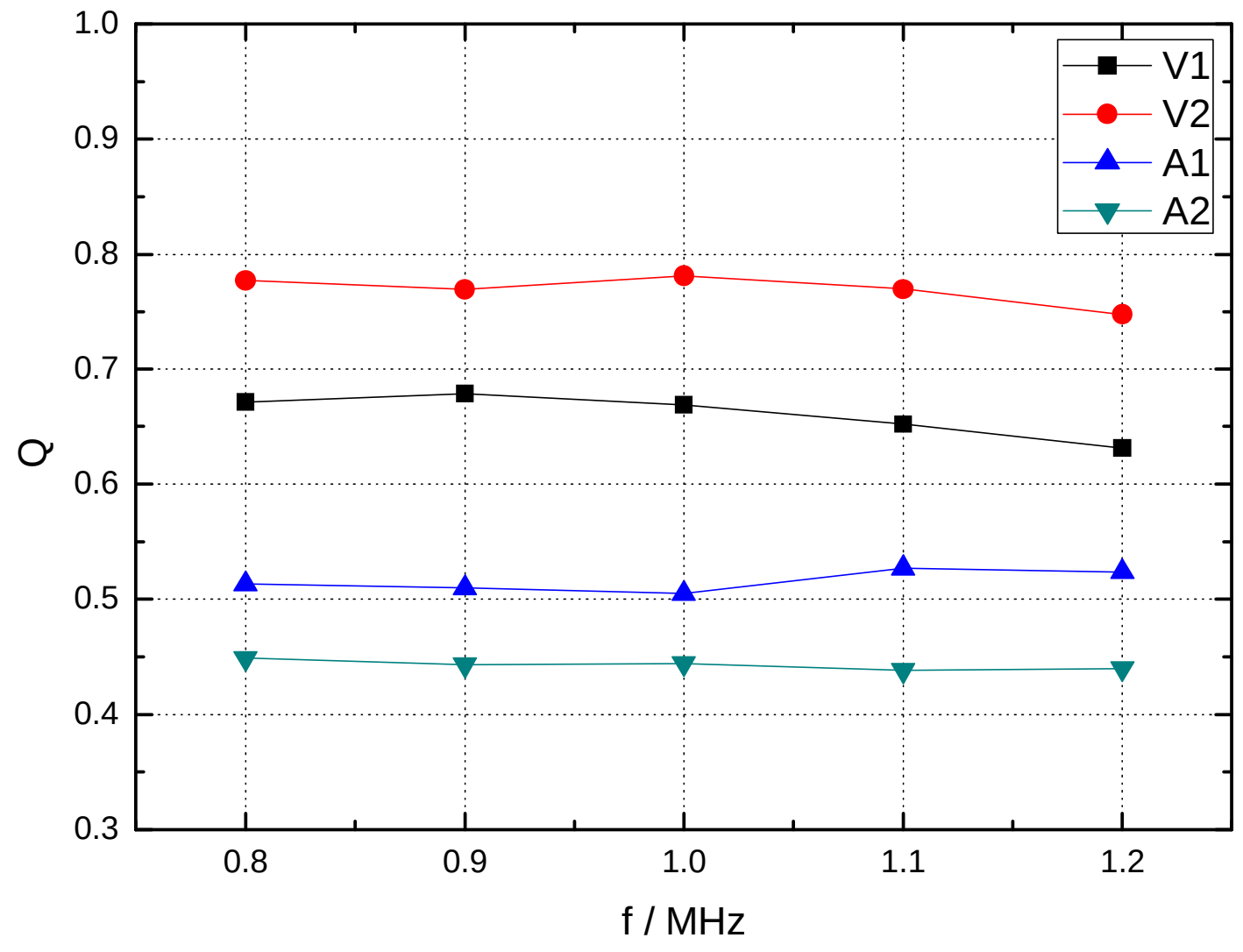

Fig.3. Frequency dependence of Q value of V1, V2, A1 and A2

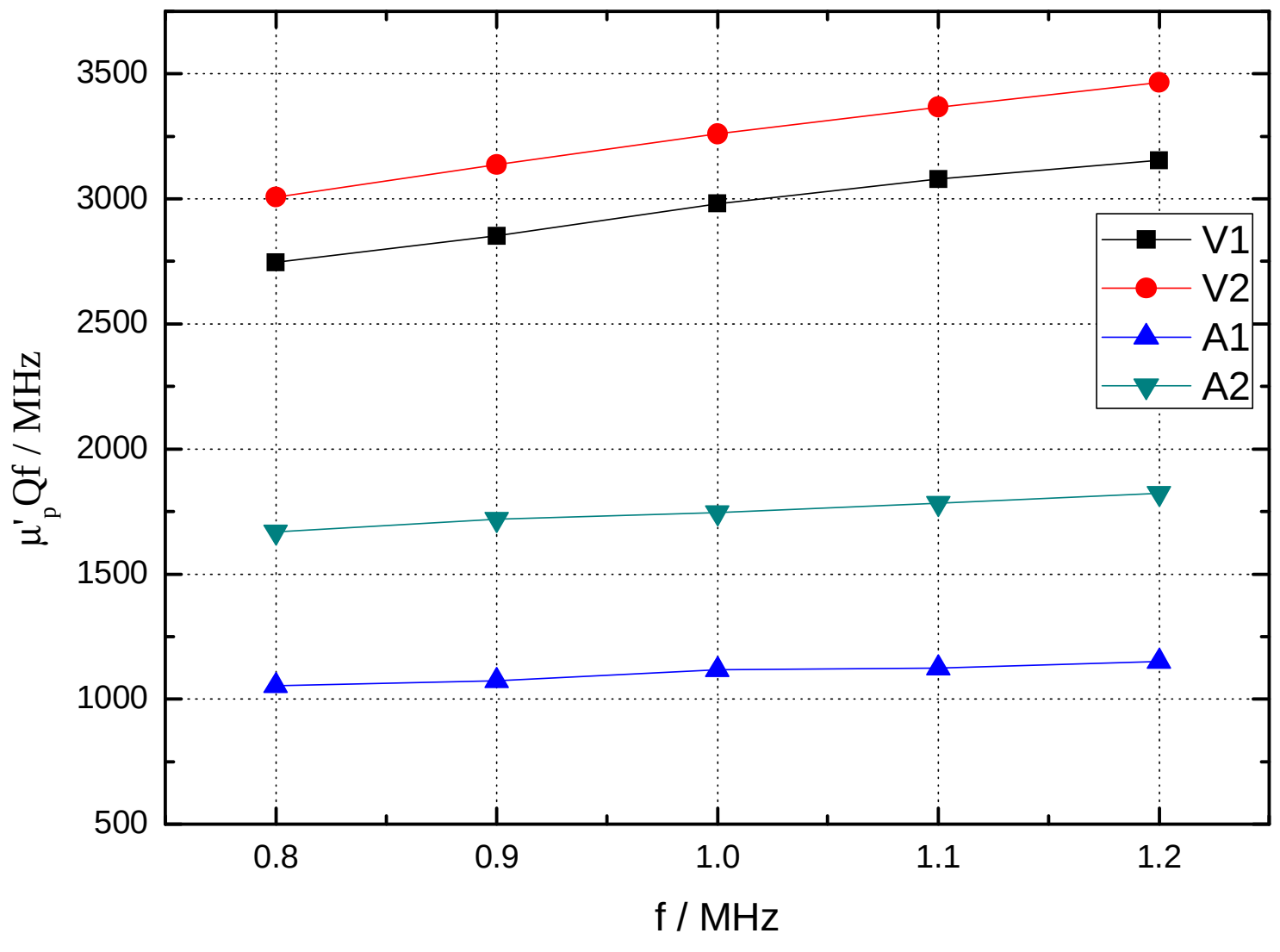

Fig.4. Frequency dependence of $\mu_p$'Qf product of V1, V2, A1 and A2

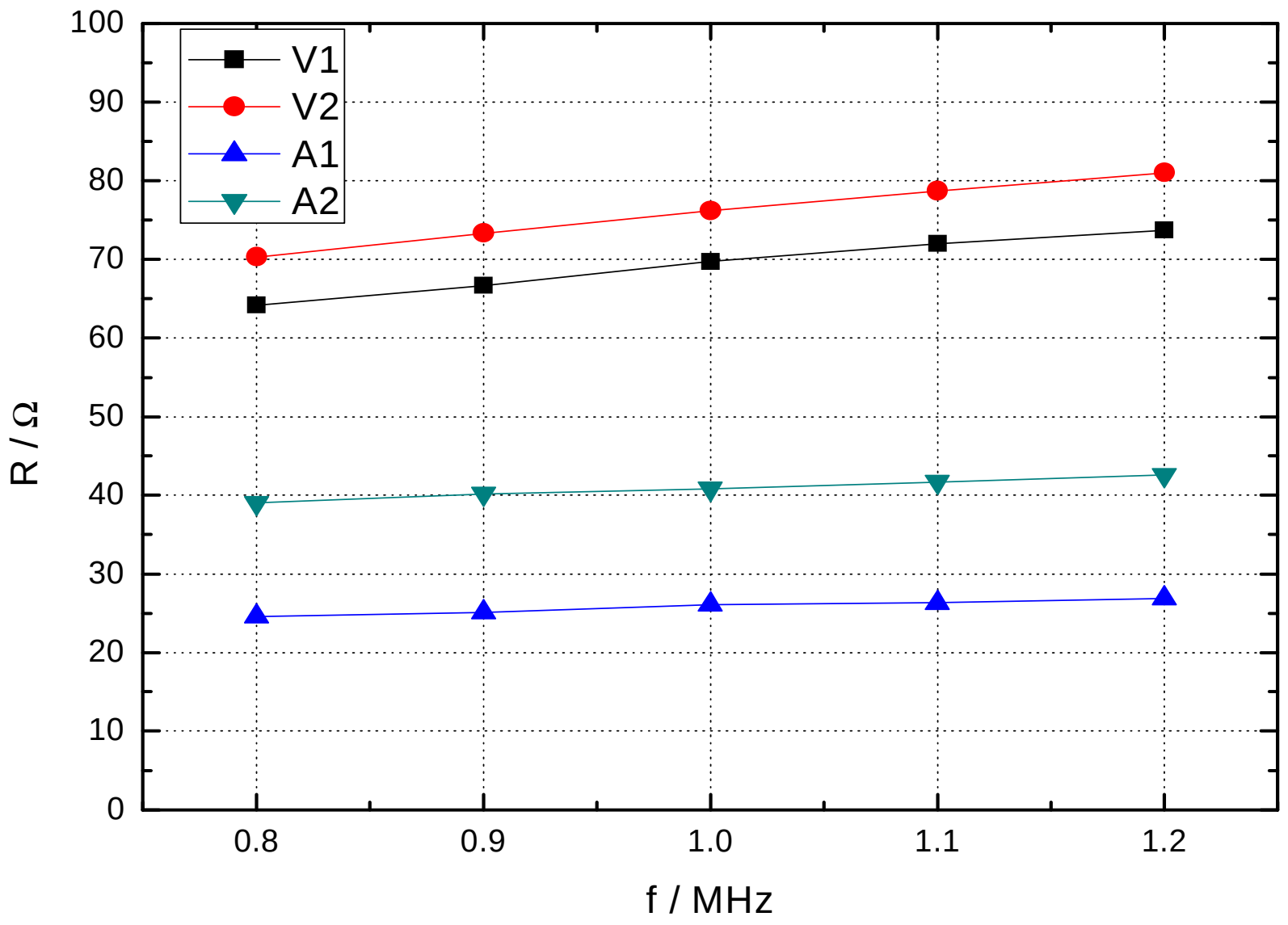

Fig.5. Frequency dependence of R value of V1, V2, A1 and A2

From Fig.3 we could see the Q value of the type V1 is about 0.67 at 1.0MHz, the change is very small when frequency change from 0.8 MHz to 1.2MHz. The Q value of type V2, A2 and A2 are about 0.78, 0.51 and 0.44 at 1.0MHz respectively. And the change is very small too.

From Fig.4 we see that, the $\mu_p$'Qf value of the type V1 is about 2.98GHz at 1.0MHz, the change from 2.75GHz to 3.15GHz when frequency change from 0.8 MHz to 1.2MHz. The $\mu_p$'Qf value of type V2, A2 and A2 are about 3.26GHz, 1.12 GHz and 1.75 GHz at 1.0MHz respectively. The corresponding variation range is from 3GHz to 3.46GHz, 1.05GHz to 1.15GHz and 1.67GHz to 1.82GHz respectively when frequency varied from 0.8MHz to 1.2MHz.

From Fig.5 we see that, the shunt impedance R of the type V1 changes from 64.2Ω to 73.7Ω when frequency varied from 0.8 MHz to 1.2MHz. The R of type V2, A2 and A2 change from 70.3Ω to 81Ω, 24.6Ω to 26.9Ω and 39Ω to 42.6Ω respectively.

Therefore, we could conclude that the type V2 and A2 have higher Q value, $\mu_p$'Qf and R in the same type material. For the same type material the property of the core which has the insulation layer of silica gel is better than the insulation layer of SiO2 and silica gel.

Finally, in our practical application we selected the cores with outer radius of 400mm, inner radius of 170mm and thickness of 30mm. If the characteristics of the MA core do not change when scale up the dimension of the core, we would calculate the shunt impedance of single core with the dimension of practical application. So, we calculate the shunt impedance of V1, V2, A1 and V2 in the frequency range from 0.8 to 1.2MHz. The calculation results are shown in Fig.6.

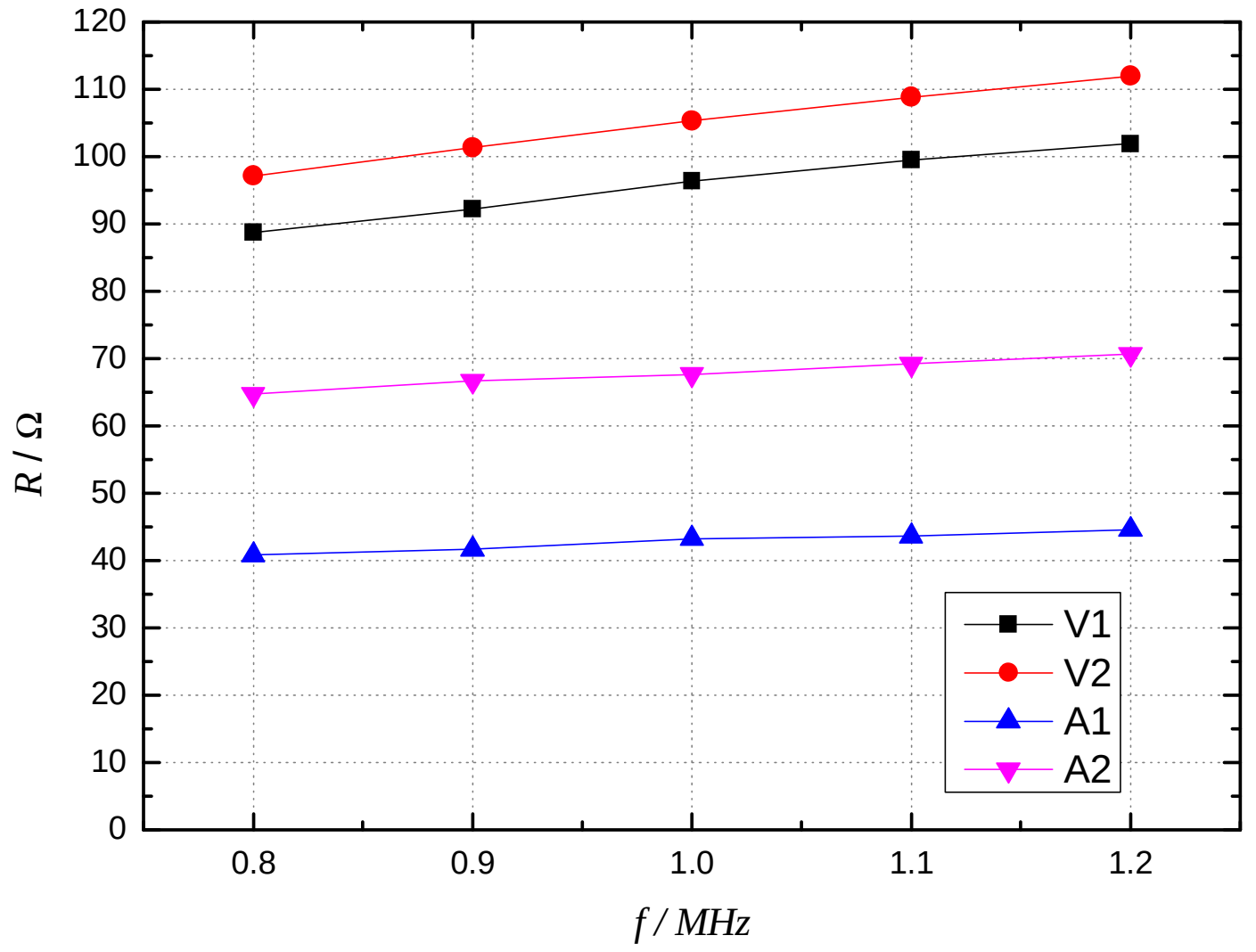

Fig.6. Frequency dependence of R value of V1, V2, A1 and A2 with practical dimension.

From Fig.6 we could see type V2 has the highest shunt impedance, it will be the best selection to load RF cavity in four kinds of material. And type V has the higher shunt impedance than type A.

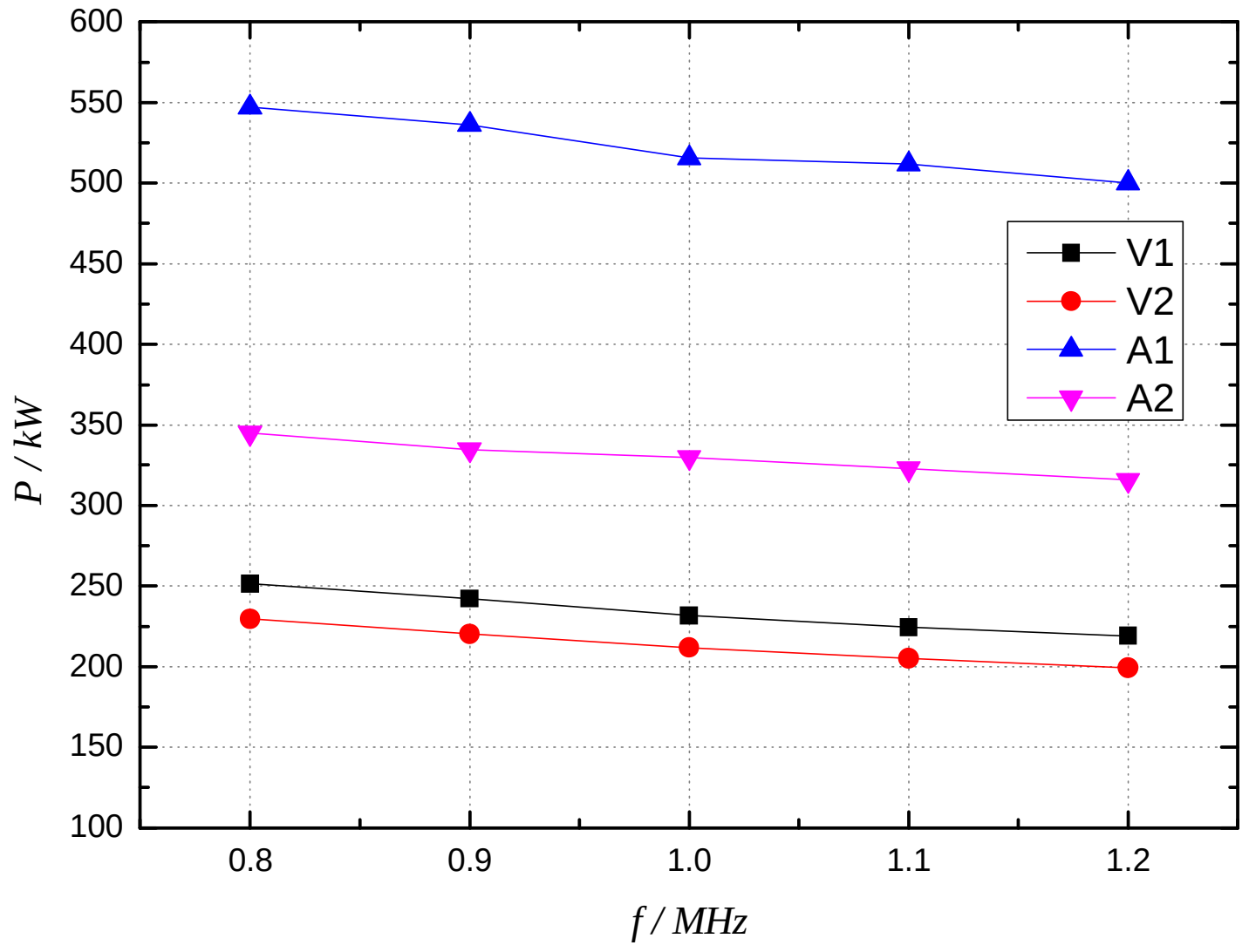

Fig.7. Frequency dependence of power of V1, V2, A1 and A2 with practical dimension.

Furthermore we calculated the requirement power when provided 50kV accelerating voltage by using above material. And the results are shown in Fig.7. In our cavity design we needed 14 cores to load RF cavity [3]. From the diagram we see that, in order to obtain 50kV accelerating voltage the power source must provide 230kW power in the frequency range from 0.8 to 1.2MHz when we choose type V2 to load RF cavity, or 250kW if we chose V1. While 350kW needed if we choose type A2, 550kW needed if we choose type A1.

## 4. Conclusion

According to the measurement for the material which produced by Liyuan Company, some characteristics of the type V1, V2, A1 and A2 have been obtained. In order to get 50kV accelerating voltage when provided limited power source, the type V1, V2 and A2 could be as candidate material to load the RF cavity. The best case is to choose type V2 to load RF cavity because the shunt impedance of single core is larger than the others, so the power requirement will be lower than the others. But we must take into account the price of different material, if the price of type V is more expensive than type A, the type A2 will be a compromising selection.